# The Magellanic Clouds Survey: a Bridge to Nearby Galaxies

Paul Scowen

School of Earth & Space Exploration
Arizona State University
PO Box 871404, Tempe, AZ 85287-1404

(480) 965-0938

paul.scowen@asu.edu

Rolf Jansen (Arizona State University)

Matthew Beasley (University of Colorado – Boulder)

Daniela Calzetti (University of Massachusetts)

Alex Fullerton (STScI)

John Gallagher (University of Wisconsin – Madison)

Mark McCaughrean (University of Exeter)

Robert O'Connell (University of Virginia)

Sally Oey (University of Michigan)

Nathan Smith (University of California – Berkeley)

**Abstract**

We outline to the community the value of a Magellanic Clouds Survey that consists of three components: I) a complete-area, high resolution, multi-band UV-near-IR broadband survey; II) a narrowband survey in 7 key nebular filters to cover a statistically significant sample of representative HII regions and a large-area, contiguous survey of the diffuse, warm ISM; and III) a comprehensive FUV spectroscopic survey of 1300 early-type stars. The science areas enabled by such a dataset are as follows:  A) assessment of massive star feedback in both HII regions and the diffuse, warm ISM; B) completion of a comprehensive study of the 30 Doradus giant extragalactic HII region (GEHR); C) development and quantitative parameterization of stellar clustering properties; D) extensive FUV studies of early-type stellar atmospheres and their energy distributions; and E) similarly extensive FUV absorption-line studies of molecular cloud structure and ISM extinction properties.  These data will also allow a number of additional studies relating to the underlying stellar populations.

**Introduction & Scientific Context**

This science program consists of a high spatial resolution wide field imaging survey across the Magellanic Clouds. The central goal of this program is to apply knowledge derived from local star forming environments to analogous regions where we can still resolve important physical scales to relate local star formation properties to  those processes operating on global scales. The study will be extended to regions that do not have nearby analogs but are common in other external galaxies. This systematic, hierarchical approach provides for the first time a statistically supported application of local star formation knowledge to environments and conditions that do not have analogs in our own Galaxy.  By extending the study to nearby galaxies we provide access to, and can study the effects of, more extreme radiation environments, lower metallicities, superbubble boundaries, and so on.  We can use results from local surveys to infer for the first time the characteristics of low-mass star-forming environments in the Magellanic Clouds.  Bearing in mind the other missions that will soon be active, we will aim to also use complementary IR observations from the ground or using missions such as JWST to study the multi-wavelength properties of star forming environments, the rate of subclustering, small and large-scale feedback effects, and the propagation of star formation over small and large spatial scales. Such a survey will require an efficient, wide-field (10s of arcmin on a side) camera on a large (~2-4 m) aperture space telescope, which will be able to map both Magellanic Clouds in their entirety at < 0.1" resolution in multiple mid-UV (~200nm)-near-IR (Y-band) broadband filters to $m_{AB} > 26$ mag and in key diagnostic narrowband filters to $10^{-16}$ erg cm$^{-2}$ s$^{-1}$ arcsec$^{-2}$.

**Compelling Science Themes Based on Recent Advances**

**Feedback from Massive Stars**

Massive OB stars have a profound influence on their environment, ranging from destructive evaporation of molecular clouds that curtails further star formation, to galactic-scale production of ionizing radiation, galactic superwinds, and heavy elements that drive evolutionary processes in galaxies and the cosmos itself.  The Magellanic Clouds, owing to their proximity and minimal Galactic obscuration, are a superior test-bed in which to examine both triggering and feedback processes on both microscopic and macroscopic scales.

With such a deep, narrowband imaging survey of the Clouds, the elusive large-scale ionization structure of the diffuse, warm ionized medium (DWIM) will become dramatically more apparent, allowing its spatial and ionization properties to be readily correlated with embedded star-forming regions, which presently are presumed to be the origin of the DWIM. This survey will also offer important leverage on the DWIM properties with respect to 3-D ISM structure and metallicity between the LMC and SMC.

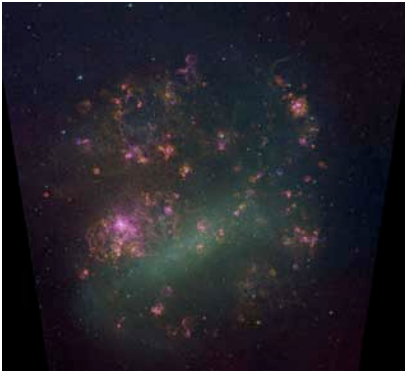

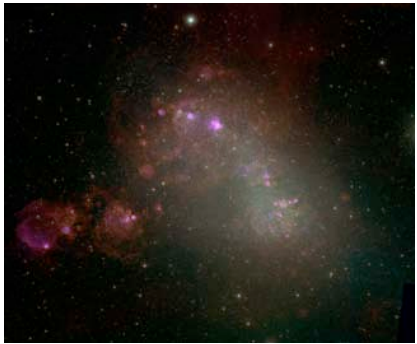

**Figure 1**: MCELS Mosaics of the Magellanic Clouds – this survey would offer the same coverage but at a much higher resolution and greater depth to answer a whole new set of scientific questions.

Comparison with the quantitative field massive star populations, as well as those in OB associations (see below) will provide unprecedented constraints on feedback parameters such as ionizing fluxes, stellar wind power, and elemental enrichment. It will be possible, with this survey dataset, to quantify and parameterize the spheres of influence of massive stars as a function of mass and interstellar conditions, for the three feedback effects.

Intermediate to the large and small-scale feedback effects is the transition stage corresponding to superbubbles and supernova remnants. How energy and mass are transferred to the surrounding ISM remains to be clearly understood. Even the crude energy budgets of superbubbles are currently sketchy, with uncertainty as to whether their evolution is purely adiabatic. This survey map of the Clouds will yield high-resolution data for dozens of superbubbles and SNRs, with which the shock structures, parent stellar populations, and ionization properties can be examined in unambiguous detail. Combined with the wealth of multi-wavelength survey data (MCELS, Spitzer SAGE survey) now available for the Clouds, we will finally have an opportunity for a major breakthrough in understanding the transfer of energy from massive stars to the diffuse ISM.

**30 Doradus: The Nearest Giant Extragalactic HII Region**

30 Doradus is a unique, giant star formation complex in the Large Magellanic Cloud (LMC). At a distance of 51.3 kpc, 1" corresponds to a linear scale of ~0.25 pc and a 0.01" pixel would sample ~0.0025 pc or ~500 AU. The nebula is centered on a dense, massive cluster of newly formed stars, the densest component of which is called R136. The nebula itself is more than 180 pc across, which qualifies it as a smaller member of the elite class of giant extragalactic HII regions. If 30 Doradus were placed at the distance of the Orion Nebula from Earth, it would appear to be more than 20° across and would fill more than 4% of the night sky.

R136 is very compact and contains several hundred OB stars with a number of Wolf-Rayet stars. The integrated UV flux from this cluster is intense: more than 50 times that being produced in the center of the Orion Nebula. It has been shown that the majority of the stars in the cluster were formed in a single star formation event ≈2-3 Myr ago. More than 3000 stars have been resolved in this cluster of which more than 300 are OB stars capable of producing the intense UV radiation and strong stellar winds responsible for forming and shaping the HII regions we observe in galaxies. The level of star formation exhibited by the 30 Doradus region and the neighboring LMC complex are the closest examples of starburst-modeltype star formation and as such, from our vantage point we have a unique view of the in situ environment.

The 30 Doradus Nebula plays a key role in our understanding of HII regions. In nearby regions within our own Galaxy, we can study the physical processes in detail. Work on M16 has shown that emission within the nebula arises predominantly within a narrow region at the interface between the HII region and the molecular cloud. However, an HII region like M16 is tiny in comparison with giant HII regions, and no giant HII regions are close enough to allow the stratified ionization structure of the photoevaporative flow to be studied directly. 30 Doradus alone offers an opportunity to bootstrap the physical understanding of small nearby HII regions into the context of the giant regions seen in distant galaxies. 30 Doradus is in a very different class of object from objects such as the Orion Nebula in terms of structure, size, dynamics, level of star formation, diversity of morphologies, and so on - we cannot learn what we need to know about GEHR's from the study of simple Galactic HII regions. It is critical that we learn about 30 Doradus in and of itself.

**The Clustering of Star Formation**

Because of obscuration by dust, we have no clear view and understanding of star formation on large and global scales within our Galaxy. Such understanding must come from the study of other galaxies. On large scales, perhaps the most basic concept is that of a coherence length. If star formation proceeds in pockets that are independent of one another, then there should be no correlation of ages and distances between the different regions. Alternatively, if star formation propagates in a 'wave' from one side of the galaxy to the other, then there should be a linear correlation. When a star formation 'wave' dies out on scales larger than this distances (the coherence length), then the correlation between average age and separation disappears. Ground-based studies indicate that within the LMC ages and separations between star clusters are strongly correlated up to separations of ~1°. The correlation vanishes at larger separations, perhaps because the coherence length is limited by the thickness of the LMC disk or by the Jeans length.

Hierarchical clustering can also be studied in this manner. Another fundamental parameter, hierarchical clustering, can be addressed in a similar manner with the proposed Magellanic Clouds survey. Clustering of massive stars and associations within the SMC appears to have an associated characteristic length scale of 30-60 pc. This result is very sensitive to the spatial resolution of the survey data used, underscoring the need for the high resolution of the proposed survey, which aims to quantify the, yet unknown, clustering of lower-mass stars. The characteristic clustering lengths, as a function of stellar mass, are a vital parameter for understanding the macroscopic process

The Magellanic Clouds Surveyof star formation in galaxies. A survey assembled with a wide field of view, outstanding spatial resolution and sensitivity, will provide the perfect dataset to study coherence lengths of star formation in external galaxies. Such a survey would need to image the entire LMC and SMC in several near-UV-near-IR filters to the Main Sequence turn-off for old clusters in the LMC ($m_{AB}^{MSTO} \leq 22$ mag, at $>10\sigma$) to ultimately allow measuring the ages of each cluster individually.

**Atmospheric Properties of Massive Stars**

Massive, OB-type stars dominate the return of matter and energy to the interstellar medium through their dense, fast stellar winds, intense radiation fields, and eventual death and partial dispersal in supernova explosions. Consequently, the fundamental stellar parameters of these objects must be characterized accurately, in order that their evolutionary state can be inferred and their yields of energy, momentum, and chemically-enriched feedback to the ISM can be estimated reliably. These properties are an essential input to understanding the feedback cycle for both localized molecular cloud destruction and large-scale ISM and IGM feedback effects. Since the predicted evolution of OB-type stars is a strong function of metallicity, this empirical characterization must occur for stars with different abundances so that patterns can be established and theoretical predictions can be tested.

The availability of high-quality optical and UV spectra has shown that some fundamental aspects of the atmospheres of OB-type stars are not well understood. Uncertainties in the ionizing fluxes remain at the 0.8 dex level. There is growing evidence that the winds are not smooth and homogeneous, but are instead substantially clumped. The wind properties strongly affect the ionizing outputs through wind-blanketing in the stellar atmospheres. However, in recent years it has become apparent that the loss of angular momentum through the winds also substantially affects the stellar rotation, and therefore, the entire evolution of the stars. Thus, our survey includes an FUV spectroscopic component (down to 100nm to include specific line diagnostics) to survey OB stars in the Magellanic Clouds to fully investigate their wind properties and ionizing fluxes.

A comprehensive, FUV-UV spectroscopic survey of the hottest, most massive stars in the Magellanic Clouds will provide a data set of unprecedented diagnostic power to address these issues. The Magellanic Clouds are preferable to the Galaxy for such studies, since: (a) their distances are known, hence the luminosities of individual stars can be determined reliably; (b) a range of metallicities can be probed; and (c) the UV extinction is comparatively minor. Although STIS and FUSE have observed modest but well-chosen samples of stars in the Large and Small Magellanic Clouds, this science driver advocates a distinctly new capability. In contrast to FUSE, a large aperture combined with a high-throughput FUV spectrograph will offer the spatial resolution required to observe individual OB stars in their dense natal clusters.

Such a new facility should also provide enough light-gathering power to push observations several magnitudes fainter than FUSE could reach. The type of comprehensive FUV survey of the most massive stars we have in mind will extend to spectral types of B2 V, and will include some 1300 stars. This large sample will overcome the small-number statistics that currently bias the available samples of Magellanic OB-stars with UV spectra, and will provide the first opportunity for complete

*Scowen et al.*     4

of star formation in galaxies. A survey assembled with a wide field of view, outstanding spatial resolution and sensitivity, will provide the perfect dataset to study coherence lengths of star formation in external galaxies. Such a survey would need to image the entire LMC and SMC in several near-UV-near-IR filters to the Main Sequence turn-off for old clusters in the LMC ($m_{AB}^{MSTO} \leq 22$ mag, at $>10\sigma$) to ultimately allow measuring the ages of each cluster individually.

**Atmospheric Properties of Massive Stars**

Massive, OB-type stars dominate the return of matter and energy to the interstellar medium through their dense, fast stellar winds, intense radiation fields, and eventual death and partial dispersal in supernova explosions. Consequently, the fundamental stellar parameters of these objects must be characterized accurately, in order that their evolutionary state can be inferred and their yields of energy, momentum, and chemically-enriched feedback to the ISM can be estimated reliably. These properties are an essential input to understanding the feedback cycle for both localized molecular cloud destruction and large-scale ISM and IGM feedback effects. Since the predicted evolution of OB-type stars is a strong function of metallicity, this empirical characterization must occur for stars with different abundances so that patterns can be established and theoretical predictions can be tested.

The availability of high-quality optical and UV spectra has shown that some fundamental aspects of the atmospheres of OB-type stars are not well understood. Uncertainties in the ionizing fluxes remain at the 0.8 dex level. There is growing evidence that the winds are not smooth and homogeneous, but are instead substantially clumped. The wind properties strongly affect the ionizing outputs through wind-blanketing in the stellar atmospheres. However, in recent years it has become apparent that the loss of angular momentum through the winds also substantially affects the stellar rotation, and therefore, the entire evolution of the stars. Thus, our survey includes an FUV spectroscopic component (down to 100nm to include specific line diagnostics) to survey OB stars in the Magellanic Clouds to fully investigate their wind properties and ionizing fluxes.

A comprehensive, FUV-UV spectroscopic survey of the hottest, most massive stars in the Magellanic Clouds will provide a data set of unprecedented diagnostic power to address these issues. The Magellanic Clouds are preferable to the Galaxy for such studies, since: (a) their distances are known, hence the luminosities of individual stars can be determined reliably; (b) a range of metallicities can be probed; and (c) the UV extinction is comparatively minor. Although STIS and FUSE have observed modest but well-chosen samples of stars in the Large and Small Magellanic Clouds, this science driver advocates a distinctly new capability. In contrast to FUSE, a large aperture combined with a high-throughput FUV spectrograph will offer the spatial resolution required to observe individual OB stars in their dense natal clusters.

Such a new facility should also provide enough light-gathering power to push observations several magnitudes fainter than FUSE could reach. The type of comprehensive FUV survey of the most massive stars we have in mind will extend to spectral types of B2 V, and will include some 1300 stars. This large sample will overcome the small-number statistics that currently bias the available samples of Magellanic OB-stars with UV spectra, and will provide the first opportunity for complete characterization of the atmospheric properties of these populations. An important by-product will be the wealth of high-quality observations of interstellar absorption lines, that will be used to investigate the distribution and kinematics of hot ($10^6$ K) ionized gas, cool molecular $H_2$ gas, and dust properties along each sight line. The former will be especially important in studying mechanical feedback in individual star-forming regions, while also offering contrasting observations in the diffuse ISM.

**The Origin and Evolution of Molecular Clouds**

We furthermore need to make a comprehensive effort to probe the origins of molecular clouds by studying both FUV absorption lines ($H_2$, CO, and atomic species) and extinction by dust in the LMC and SMC. A large-aperture FUV spectrograph will far surpass FUSE as a probe of molecular cloud origins at higher extinctions and with orders of magnitude more sources than FUSE could access. The key issue is how molecular clouds, the precursors of star formation, coalesce from the diffuse ISM. This process is very poorly understood in our own Galaxy because our place in the Galactic disk prevents examination of individual clouds except at very small distances. Even if Galactic trends come to be better understood, the physical processes that lead to GMC formation may depend on their environment, differing substantially in any particular galaxy that is not a spiral with modest star formation.

The different viewing geometry and morphology of the Clouds will enable detailed examination of the early evolution of molecular clouds. FUSE has demonstrated that robust star formation (via bright UV fluxes) and low metallicity (via low dust abundance) in the Clouds combine to inhibit the formation of $H_2$ in the diffuse ISM. FUV extinction studies of hot stars in the Magellanic Clouds show evidence for a smaller population of dust grains than in our own Galaxy. This suggests that environment affects the diffuse ISM, but it remains unclear whether and how feedback influences the GMC formation, and therefore the star formation rate and efficiency. Samples of hot stars for ISM absorption-line and extinction-curve studies can be chosen directly from the wide field imaging survey described earlier, allowing us to correlate young stellar populations with the properties of the nearby ISM gas and dust and to address the question of how GMCs arise and how star formation regulates itself on galactic scales. This ambitious goal requires both accurate photometry for the stellar populations (to derive reddening for the candidate stars) and a high-throughput FUV spectrograph.

These data will extend the excellent work done in the MCs by FUSE to new "extragalactic galactic" environments. One of the legacies of the FUSE mission is its detailed study of the interstellar medium of the Small and Large Magellanic Clouds, using roughly 200 observations of OB stars. With as much as 25 times the effective collecting area, FUSE's work can be extended to higher extinction at twice the resolution. At low extinctions $H_2$ in the Clouds responds to the environmental influences in the sense that the $H_2$ abundance (relative to H) is reduced at low metallicity while its rotational excitation is enhanced by FUV radiation from nearby hot stars. Using fainter sources than are accessible to FUSE, these data will extend this work above $E(B-V) = 0.3$, for correlations of $H_2$ abundance and excitation with local conditions. Understanding such correlations is crucial for the interpretation of the properties of gas at higher redshifts (starburst, DLAs, etc.). This project will require complete FUV spectra of 10 - 20 stars.

**Key Advances in Observation Needed**

To achieve the science goals of this program a variety of capabilities need to be implemented. The majority of the tracers and the various phases of the ISM and stellar populations being targeted require the angular resolution and wavelength agility of a medium to large aperture (2-4m) UV/optical space telescope combined with a wide-field imaging camera that can provide diffraction-limited images into the UV-blue to capture the UV-bright stellar populations that HST has been unable to reach. Such a telescope needs to be located in an orbit that is both dynamically and thermally stable (such as L2) to produce the photometric stability required by many aspects of the science goals. A broad complement of both broad- and narrow-band filters will be necessary to isolate and measure not only the unique tracers of specific atomic species but also the trends in stellar color across entire swaths of our local Galactic neighborhood.

Over the next decade the specific technological capabilities that need to be developed include the ability to construct large focal plane arrays that are flight rated for space in a reliable and straightforward fashion that simultaneously mitigates the risk, maximizes the yield rate and keeps the costs down. This is a major challenge that affects not only this project but many others, and requires real investment on the part of the community to allow such systems to be built routinely. In addition, the design of next generation coatings and dichroic optical elements will allow for the design and construction of truly advanced telescope/camera systems that can yield remarkable advances in imaging efficiency for a minimal investment.

This project requires the combination of the wide field, high resolution imager with a high-throughput FUV spectrograph capable of detecting down to 1000Å to provide access to key diagnostic lines. Such a capability will require the development of reliable coatings that are highly reflective in the FUV and that can be applied to larger optics than has been achieved so far. Next generation UV detectors provide the efficient detection of weak signals will also be necessary to make this capability efficient and reliable.

**Four Central Questions to be Addressed**

1. What is the nature of the interrelation between the formation, evolution and destruction of massive stars and the energization of the DWIM? How does the formation of massive stars in a particular locale affect and dictate the subsequent star formation across that region?
2. What is the fundamental difference between starburst star formation and the more common disk modes we see in disk star forming regions in our own Galaxy? What causes the several orders of magnitude increase in star formation efficiency as well as the almost instantaneous formation of thousands of stars at once?
3. What is the correct density and velocity structure associated with the stellar winds from massive stars? How does inhomogeneity and clumping in these winds affect the transfer of energy and material to the ISM and the process of recycling of material from the stellar to the gas phase for the next generation of stars?
4. What are the global processes that govern the assembly and evolution of giant molecular clouds? Since these nurseries host the most dominant modes of star formation in galaxies, we need to understand the nature of their formation and development if we are to understand the underlying process of stellar assembly.

**Area of Unusual Discovery Potential for the Next Decade**

While the science program in this paper have defined a loose set of specifications (see Table 1), it should be recognized that the opportunity for truly unique discovery is made possible by the **combination** of both a wide angular field of view (tens of arcminutes on a side) **with** the diffraction limit of a medium to large aperture in the UV/optical (resolution elements below 10-20 mas). HST and JWST have provided and will provide exquisite resolution but over very small fields of view. Many problems, such as those discussed here, and others such as the nature of the Universe around the time of Reionization, require not only large collecting area and high resolution, but large fields of view to locate and measure very rare objects, or suites of objects whose location cannot be known a priori. The potential discovery rate from such a combined capability cannot be underestimated, and should be very seriously considered by the Decadal Survey.

| Parameter | Specification | Justification |
|---|---|---|
| Field of View | At least 200 sq. arcmin | To allow a statistically complete survey of as many targets and environments as possible in a reasonable period of time |
| Resolution | Diffraction Limited to 300nm | To provide access to UV-blue stellar populations; to resolve structure in ionization fronts, etc. |
| Aperture | 1.5-4m | This is driven by the limiting magnitudes needed traded against the necessary exposure times to achieve them – the larger the better |
| Stability | A small percentage of a pixel | To allow the stable photometry and astrometric measurements necessary to achieve the science goals |
| Photometric Stability | Combination of gain, A/D conversion and QE need to be stable to better than $10^{-5}$ | Again to provide the photometric stability to achieve the science goals of the project |
| Filter Suite | F250W, F336W, F438W, F625W, F775W, F850W; F547M, F980M, F1020M, F1050M, F1080M; F280N, F373N, F469N, F487N, F502N, F631N, F656N, F673N, F953N | Dictated by both broad-band colors needed to survey stellar populations and the narrow-band diagnostics necessary to probe the resolved gas structure and dynamics |
| Optical Design | Efficient design offering a wide, well-corrected field of view to be populated by a large focal plane | The science program can only be achieved by an efficient design that offers parallel observing in the red and blue, with little field distortion, and as large an objective as possible |
| Detectors | High yield, efficient detectors, customized in their response to the passbands needed | Tiling the large focal plane will be challenging – we need an efficient manufacture and testing process, combined with the ability to match response to the optical channels |
| Coatings | Development of stable, high-reflectivity FUV mirror coatings | To provide high throughput access to the FUV (below 115nm) while minimizing risk to the optical reflectivity of an optical system |
| FUV Detectors | Development of next generation MCP technology | To provide a low-cost, high QE, robust solution to allow efficient observations of FUV emission, below 115nm to as low as 100nm |
| FUV Spectroscopic Resolution | R > 30,000 | To enable sufficient resolution to see structure and dynamics of emission from science targets |

**Table 1**: Science Driven General Specifications